\documentstyle[aps,prl]{revtex}
  \advance\topmargin      by -0.5in
\begin{document}
\draft
\title{
Phase separation in systems with charge ordering}
\author{
  M.Yu. Kagan $^{(a),(b)}$,
  K.I. Kugel $^{(c)}$
  and D.I. Khomskii $^{(d)}$}
\address{$^{(a)}$  P.L. Kapitza Institute for Physical Problems,
Russian Academy of Sciences, Kosygina Str.\ 2,
117334 Moscow, Russia\\ $^{(b)}$ Max-Planck-Institut f\"{u}r Physik
Komplexer Systeme, N\"{o}thnitzer Str.\ 38, D-01187 Dresden, Germany\\
$^{(c)}$ Institute of
Theoretical and Applied Electrodynamics, Russian Academy of Sciences,
Izhorskaya Str.\ 13/19, \\127412 Moscow, Russia\\
$^{(d)}$ Laboratory of Applied and Solid State Physics, Materials
Science Center, University of Groningen, Nijenborgh 4,\\
9747AG Groningen, The Netherlands}
\maketitle
\widetext


\begin{abstract}
\begin{center}
\parbox{14cm}
{A simple model of charge ordering is considered. It is
shown explicitly that at any deviation from half-filling
($n \neq 1/2$) the system is unstable with respect to phase
separation into charge ordered regions with $n = 1/2$ and
metallic regions with smaller electron or hole density.
Possible structure of this phase-separated state (metallic droplets in a
charge-ordered matrix) is discussed.
The model is extended to account for the strong
Hund-rule onsite coupling and the weaker intersite
antiferromagnetic exchange. An analysis of this extended
model allows us to determine the magnetic structure of the
phase-separated state and to reveal the characteristic features
of manganites and other substances with charge ordering.}
\end{center}
\end{abstract}
\pacs{
\hspace{1cm}
}
\section{Introduction}
The problem of charge ordering (CO) in magnetic oxides attracts an
attention of theorists since the discovery of Verwey transition in
magnetite in the end of thirties \cite{Verwey}. An early
theoretical description of this phenomenon was given e.g. in
\cite{Khomskii0}. Recently this problem was reexamined in a number
of papers in connection with the colossal magnetoresistance in
manganites, see e.g. \cite{Mutou,Khaliullin,Jackeli}. The
mechanisms stabilizing the CO state may be different: the Coulomb
repulsion of charge carriers (the energy minimization requires
keeping the carriers as far away as possible, similar to the
Wigner crystallization), or the electron-lattice interaction
leading to the effective repulsion of electrons at the nearest
neighbor sites. In all cases, charge ordering can arise in systems
with the mixed valence if the electron bandwidth is sufficiently
small --- large electron kinetic energy stabilizes the homogeneous
metallic state. In real materials, in contrast to the Wigner
crystallization, the underlying lattice periodicity determines the
preferential types of CO. Thus, in the simplest bipartite lattice,
to which belong the colossal magnetoresistance manganites of the
type R$_{1-x}$A$_{x}$MnO$_{3}$ (R = La, Pr; A = Ca, Sr), or
layered manganites R$_{2-x}$A$_{x}$MnO$_{4}$,
R$_{2-2x}$A$_{1+2x}$Mn$_{2}$O$_{7}$, the optimum conditions for the
formation of the CO state exist for doping $x = 1/2$. At such
value of $x$ the concentrations of Mn$^{3+}$ and Mn$^{4+}$ are
equal, and the simple checkerboard arrangement is possible. The
most remarkable experimental fact here is that even at $x \neq
1/2$ (in the underdoped manganites, $x<1/2$) only the simplest
version of charge ordering is experimentally observed with
alternating checkerboard structure of occupied and empty sites in
the basal plane \cite{Jirak}. In other words, this structure
corresponds to the doubling of the unit cell, whereas the more
complicated structures with longer period (or even incommensurate
structures) do not actually appear in this case.

Then, the natural question arises: how could we redistribute the
extra or  missing electrons in the case of arbitrary doping level,
keeping the superstructure the same as for $x=1/2$~? To answer this
question, the experimentalists introduced the concept of incipient
charge ordered state corresponding to the distortion of long-range CO
by microscopic metallic clusters \cite{Arulraj}. In fact, the
existence of such a state implies a kind of phase separation. Note
that the phase separation scenario in manganites is very popular now
\cite{Nagaev}--\cite{Gor'kov}. Nowadays, there is a growing evidence
suggesting that an interplay between the charge ordering and the
tendency toward phase separation plays an essential role in the
physics of materials with colossal magnetoresistance.

In this paper, we consider a simple model, which allows us
to clarify the situation at arbitrary doping. We include
in this model both the Coulomb repulsion of electrons on the
neighboring sites and the magnetic interactions responsible
for the magnetic ordering of manganites.
After demonstrating the instability of the system toward phase
separation in certain ranges of doping, we also consider
the simplest form of the phase separation --- the formation of
metallic droplets in an insulating matrix, estimate parameters
of such droplets, and construct the phase diagram illustrating
the interplay between charge ordering, magnetic ordering,
and phase separation.

One has to note that the mechanism of the charge ordering
considered below (the Coulomb repulsion) is not the only one. The
electron-lattice interaction can also play an important role, see
e.g.~\cite{Yunoki}. In application to manganites, one has to take
into account also orbital and magnetic interactions
\cite{Khaliullin,Yunoki,Soloviev}. These may be important, in
particular, to explain the fact that the charge ordering in
half-doped perovskite manganites is a checkerboard one only in the
basal plane, but it is ``in-phase'' in the $\it c$- direction.
However, the nature of such charge ordering is not clear yet and
presents a separate problem: it is not evident that the magnetic
interactions responsible for this stacking of $\it ab$-planes in
\cite{Yunoki} is indeed the dominant mechanism. Let us also
emphasize that the charge ordering in manganites is often observed
at higher temperatures than the magnetic ordering, and one has to look
for a model not relying heavily on magnetic interactions. Note
that in contrast to magnetic interactions, the Coulomb interaction
is one of the important factors always present in the systems
under consideration. Moreover, it has a universal nature and does
not depend critically on specific features of a particular system.
Consequently, our treatment can be applied also to other systems
with the charge ordering such as magnetite Fe$_{3}$O$_{4}$
\cite{Verwey}, cobaltites \cite{Moritomo1}, nickelates
\cite{Alonso}, etc.

\section{The simplest model for charge ordering}

Let us consider a simple lattice model for charge ordering:
\begin{equation}
\hat{H}=-t \sum_{<i,j>}c^+_{i} c_{j}+V\sum_{<i,j>} n_i n_j
-\mu \sum_i n_i,
\end{equation}
where $t$ is the hopping integral, $V$ is the nearest neighbor
Coulomb interaction (similar $nn$ repulsion can be also obtained
via the interaction with the breathing-type optical phonons);
$\mu$ is the chemical potential, and $c^+_{i}$ and $c_{j}$ are
one-electron creation and annihilation operators, $n_i=c^+_{i}
c_{i}$. Symbol $<i,j>$ denotes the summation over the
nearest-neighbor sites. Here, we omit for simplicity spin indices.
We also assume the absence of a double occupancy in this model due
to the strong onsite repulsion between electrons.

The models of the type (1) with the $nn$ repulsion being
responsible for the charge ordering are the most popular ones to
describe this phenomenon, see
e.g.~\cite{Khomskii0,Mutou,Jackeli,Bulla}  and references therein.
Hamiltonian~(1) captures the main physical effects; if necessary,
one can add to it some extra terms, which we will also do in
Section~V below.

In the main part of our
paper, we will always speak about
electrons. However in application to real manganites we
will mostly have in mind less than half-doped (underdoped)
systems of the type R$_{1-x}$A$_{x}$MnO$_{3}$ with $x < 1/2$.
Thus, for a real system one has
to substitute {\it holes} for our {\it electrons}. All the
theoretical treatment definitely remains the same (from the
very beginning we could define operators $c$ and $c^+$ in
(1) as the operators of holes); we hope that it will not
lead to any misunderstanding.

We consider below the simplest case of square (2D) or cubic (3D)
lattices, where for $x = 1/2$ the simple two-sublattice ordering
would take place. As mentioned in the Introduction, this is the
case in layered manganites,whereas the ordering in 3D perovskite
manganites is like this only in the basal plane, the ordering
being ``in-phase'' in the {\it c} direction. To account for this
behavior, apparently a more complicated model would be necessary.

For the case $n = 1/2$, the model (1) was analyzed in many papers;
we follow the treatment of the Ref.~\cite{Khomskii0}. As
mentioned above, the Coulomb
repulsion (second term in (1)) stabilizes charge ordering
in the form of checkerboard arrangement of occupied and empty
sites, whereas the first term (band energy) opposes this
tendency. At arbitrary values of electron density $n$, we
shall at first consider a homogeneous CO solution and use
the same ansatz as in \cite{Khomskii0}, namely
\begin{equation}
n_i=n[1+(-1)^i \tau]. \end{equation}

Such an expression implies the doubling of lattice
periodicity, with the
local densities $n_1=n(1+\tau)$ and
$n_2=n(1-\tau)$ at the neighboring sites. Note that at $n = 1/2$
for a general form of electron dispersion without
nesting, the CO state exists only at
sufficiently strong repulsion $V > 2t$ \cite{Khomskii0}.
The order parameter is $\tau < 1$ for finite $V/2t$, and the
ordering in general is not complete, i.e. an average
electron density $n_i$ differs from zero or one even
at $T = 0$.

We use the same coupled Green function approach
as in \cite{Khomskii0}, which yields
\begin{equation}
\left\{
\begin{array}{ccc}
  (E+\mu) G_1 -t_k G_2 -zVn(1-\tau) G_1
= {\mbox {\Large $\frac{1}{2\pi}$}}&&\\
  &&\\
  (E+\mu) G_2 -t_k G_1 -zVn(1+\tau) G_2 = 0&&
  \end{array} \right.
\end{equation}
where $G_1$ and $G_2$ are the Fourier transforms of the
normal lattice Green functions $G_{il}=\langle \langle c_i
c^+_l \rangle \rangle$ for sites $i$ and $l$ belonging
respectively to one or different sublattices, $z$ is the
number of nearest neighbors and $t_k$ is the Fourier
transform of a hopping matrix element. While deriving (3),
we performed a mean-field decoupling and replaced the
averages $\langle c^+_i c_i \rangle$ by the
onsite densities $n_i=n[1+(-1)^i \tau]$. Solution of (3)
leads to the following spectrum:
\begin{equation}
E+\mu=Vnz\pm \sqrt{(Vn\tau z)^2+t_k^2}=Vnz\pm \omega_k.
\end{equation}

The spectrum defined by (4) resembles the spectrum
of superconductor and hence the first term under the
square root is analogous to the
superconducting gap squared. In other words, we can
introduce the charge-ordering gap by the formula
$$
\Delta=Vn\tau z.
$$
It depends upon density not only explicitly, but
also via the density dependence of $\tau$.

Hence, we get
\begin{equation}
\omega_k=\sqrt{\Delta^2+t_k^2}.
\end{equation}
Note that there is one substantial difference between the
spectrum of charge ordered state (5) and superconducting
state, namely here for $n \neq 1/2$ the chemical potential does
not enter under the square root in (5) in contrast to the
spectrum of superconductor where
$$
\omega_k = \sqrt{(t_k-\mu)^2+\Delta^2}.
$$
Then, we can find the following expressions for
the Green functions :
\begin{equation}
\left\{
\begin{array}{ccc}
  &&G_1 = {\mbox {\Large $\frac{A_k}{E+\mu-Vnz-\omega_k+i0}
+\frac{B_k}{E+\mu-Vnz+\omega_k+i0}$}}\\
  &&\\
  &&G_2 = {\mbox {\Large $\frac{t_k}{2\omega_k} \frac{1}{2\pi}
\left[
\frac{1}{E+\mu-Vnz-\omega_k+i0}-\frac{1}{E+\mu-Vnz+\omega_k+i0}\right
]
$}},
  \end{array}
\right. \end{equation}
where
\begin{equation}
{\mbox {$A_k=$}}\frac{1}{4\pi} \left({\mbox
{$1-$}}\frac{\Delta}{\omega_k}\right), {\mbox ~} {\mbox { $B_k=$}}
\frac{1}{4\pi}\left({\mbox {$1+$}} \frac{\Delta}{\omega_k}\right).
\end{equation}

After the standard Wick transformation
$E+i0 \rightarrow iE$ in the expression for $G_1$,
we can find the densities in the following form
\begin{eqnarray}
&&n_1=n(1+\tau)=\int\left[ \left(1-\frac{\Delta}{\omega_k}
\right) f_F (\omega_k - \mu+Vnz )+
\left(1+\frac{\Delta}{\omega_k} \right) f_F (-\omega_k -
\mu+Vnz ) \right] \frac{d^3 {\mbox{\bf k}}}{2\Omega_{BZ}}\\
&&n_2=n(1-\tau)=\int\left[ \left(1+\frac{\Delta}{\omega_k}
\right) f_F (\omega_k - \mu+Vnz )+
\left(1-\frac{\Delta}{\omega_k} \right) f_F (-\omega_k -
\mu+Vnz ) \right] \frac{d^3 {\mbox{\bf
k}}}{2\Omega_{BZ}},\nonumber
\end{eqnarray}
where $f_F(y)=1/\left(e^{y/T} +1\right)$ is the Fermi
distribution function, and $\Omega_{BZ}$ is the volume of
the first Brillouin zone.

Summing up and subtracting two equations for
$n_1$ and $n_2$, we get the resulting system of
equations for $n$ and $\mu$:
\begin{equation}
\left\{
\begin{array}{ccc}
  1 = Vz {\mbox {\Large $\int\frac{1}{\omega_k}$}} \left[f_F (-\omega_k
-
\mu+Vnz )
  -f_F (\omega_k - \mu+Vnz )\right] {\mbox {\Large $\frac{d^3 {{\bf
k}}}{2\Omega_{BZ}}$}}&&\\
  &&\\
n ={\mbox {\LARGE $\int$}}  \left[f_F (-\omega_k - \mu+Vnz )
  +f_F (\omega_k - \mu+Vnz )\right] {\mbox {\Large $\frac{d^3 {{\bf
k}}}{2\Omega_{BZ}}$}}.&&
  \end{array}
\right. \end{equation}

For low temperatures $T\rightarrow 0$  and $n\leq \frac{1}{2}$
it is reasonable to assume that $\mu-Vnz$ is negative.
Hence $f_F (\omega_k - \mu+Vnz )=0$ and
$f_F (-\omega_k - \mu+Vnz )=\theta(-\omega_k - \mu+Vnz )$
is the step function.

It is easy to see that for $n = \frac{1}{2}$ the system of
equations (9) yields identical results for all $-\Delta
\leq \mu-Vnz \leq \Delta$. From this
point of view, $n=1/2$ is a point of indifferent
equilibrium. For infinitely small deviations from $n=1/2$,
that is, for densities $n=1/2-0$, the chemical potential
should be defined as $\mu= -\Delta + V z /2 = V z/2 \cdot (1-\tau)$.
If we consider a strong coupling case $V\gg 2t$ and assume a
constant density of states inside the band, then for a
simple cubic lattice we have $\tau=1-\frac{2W^2}{3V^2 z^2}$, and
hence
\begin{equation}
\mu =\frac{W^2}{3Vz}, \end{equation}
where $W=2zt$ is the bandwidth. Note that for the density $n = 1/2$
a charge-ordering gap $\Delta$ appears for an arbitrary
interaction strength $V$. This is due to the existence of
nesting in our simple model.  In the weak coupling case
$V \ll 2 t$  and with the perfect nesting, we have
$\Delta\sim W \exp\left\{  - \frac{W}{V z}\right\}$ and $\tau$
is exponentially small.  For $zV \gg W$ or accordingly for $V \gg 2 t$:
$\Delta \approx Vz/2$ and $\tau \rightarrow 1$.
As mentioned above, for a general form of electron dispersion
without nesting the charge ordering exists only if the
interaction strength $V$ exceeds certain critical value of
the order of bandwidth $W$ \cite{Khomskii0}.
Further on, we restrict ourselves only
to the physically more instructive case of strong coupling
$V \gg 2t$.

Now let us consider the case $n=1/2-\delta$, where $\delta\ll 1$
is a deviation from density $n=1/2$. For this case $\mu = \mu
(\delta, \tau)$, and we have two coupled equations for $\mu$ and
$\tau$. As a result,
\begin{equation}
\mu (\delta)\approx Vnz(1-\tau)-\frac{4W^2}{Vz}\delta^2 \approx
\frac{W^2}{3Vz} +
\frac{4W^2}{3Vz}\delta + O(\delta^2).
\end{equation}
Correspondingly, the energy of a charge ordered state is as follows
\begin{equation}
E_{CO}(\delta)=E_{CO}(0) - \frac{W^2}{3Vz}\delta -
\frac{2W^2}{3Vz} \delta^2 +O(\delta^3),
\end{equation}
where $E_{CO}(0)= -\frac{W^2}{6Vz}$ is the energy precisely
for density $n=1/2$ and $|E_{CO}(0)|\ll W$. At the same
time, the charge-ordering gap $\Delta$ is given by:
\begin{equation}
\Delta \approx \frac{Vz}{2}\left[
1- 2 \delta -\frac{2W^2}{3V^2 z^2}(1+4\delta) \right].
\end{equation}
Actually, the dependence of the chemical potential $\mu$
and the total energy $E$ on $\delta$, Eqs.~(11),~(12), stems
from this linear decrease of energy gap $\Delta$
with deviation from half-filling.

For $n > 1/2$ the energy of charge ordered
state starts to increase rapidly due to the large
contribution from Coulomb repulsion (the upper Verwey band
is partially filled for $n>1/2$). Contrary to the case
$n=1/2$, for $n>1/2$ each extra electron put into the
checkerboard CO state necessarily has occupied nearest
neighbor sites, increasing the total energy on $Vz|\delta|$. As a
result, we have for $|\delta|=n-1/2>0$
\begin{equation}
E_{CO}(\delta)=E_{CO}(0)+\left( Vz-\frac{W^2}{3Vz}\right) |\delta|-
\frac{2W^2}{3Vz}\delta^2+O(\delta^3).
\end{equation}

Accordingly, the chemical potential has the form
\begin{equation}
\mu(\delta) = Vz -\frac{W^2}{3Vz} -\frac{4W^2}{3Vz}|\delta|
+ o(\delta^2).
\end{equation}
It undergoes a jump equal to $Vz$ for $\tau \rightarrow 1$.
Note that the gap $\Delta$ is symmetric for $n>1/2$ and
is given by
$$
\Delta \approx \frac{Vz}{2}\left[1-2 |\delta| -
\frac{2W^2}{3V^2z^2}(1+4|\delta|) \right].
$$
We could make the whole picture symmetric with respect to $n=1/2$
by shifting all the one-electron energy levels and the chemical
potential by $Vz/2$, i.e., defining $\mu'=\mu-Vz/2$.
This change of variables, of course, would not modify our
conclusions.

\section{Phase separation}

The most remarkable implication of (11)-(15) is that the
compressibility $\kappa$ of the homogeneous charge
ordered system is negative for densities different from $1/2$,
\begin{equation}
\frac{1}{\kappa} \propto \frac{d\mu}{dn}=-\frac{d\mu}{d\delta}
=\frac{d^2 E}{d \delta^2}=-\frac{4W^2}{3Vz}<0,
\end{equation}
where $\delta=1/2-n$.
This is the manifestation of the tendency toward phase
separation characteristic of the charge ordered system with
$\delta \neq 0$. The presence of a kink in the $E_{CO}(\delta)$
(cf.~Eqs.~(12),~(14)) implies that one of the states, into which
the system might separate, would correspond to the checkerboard
CO state with $n =1/2$, whereas the other would have a certain
density $n'$ smaller or larger than 1/2. This conclusion
resembles that of \cite{Khaliullin} (see also \cite{Arovas,Kagan}),
although the detailed physical mechanism is different.
The possibility of phase separation in the model (1)
away from half-filling was also reported earlier in
\cite{Uhrig} for the infinite-dimensional case.
Below we concentrate our attention on the situation with
$n < 1/2$ (underdoped manganites); the case $n > 1/2$
apparently has certain special properties --- the existence
of stripe phases etc.~\cite{Mori}, the detailed origin
of which is not yet clear.

It easy to understand the physics of phase separation
in our case. As follows from (13), the CO gap decreases
linearly with the deviation from the half-filling.
Correspondingly, the energy of the homogeneous CO state
rapidly increases, and it is more favorable to ``extract''
extra holes from the CO state, putting them into one
part of the sample, while creating the ``pure''
checkerboard CO state in the other part of it.
The energy loss due to such redistribution of holes is
overcompensated by the gain provided by the better charge
ordering.

However, the hole-rich regions would not be completely ``empty,''
like pores (clusters of vacancies) in crystals: we can gain an
extra energy by ``dissolving'' in them a certain amount of
electrons. By doing this we decrease the band energy of the
electrons due to their delocalization. Thus, this second phase
would be a metallic one. The simplest state of this kind is a
homogeneous metal with the electron concentration $n_m$. This
concentration, as well as the relative volume of the metallic and
CO phases, can be easily calculated by minimizing the total energy
of the system. The energy of the metallic part of the sample $E_m$
is given by
\begin{equation}
E_m=-tzn_m + ct(n_m)^{5/3} + V(n_m)^2
\end{equation}
where $c$ is some constant.

Minimizing (17) with respect to $n_m$, we find the
equilibrium electron density in the metallic phase.
For the strong coupling $V > zt$, we get
\begin{equation}
n_{m0} = tz/2V
\end{equation}

Thus, in this simple treatment, the system with
$n_{m0} < n < 1/2$ would undergo phase separation
into the CO phase with $n = 1/2$ and the metallic phase
with $n = n_{m0}$. Relative volumes $v_m$ and $v_{CO}$ of
these phases for arbitrary $n$ can be found from the
Maxwell construction:
\begin{equation}
v_m/v_{CO} = (1/2-n)/(n-n_{m0}),
\end{equation}
from which we find that the metallic phase occupies
the part $v_m$ of the total volume $v$ given by the
relationship
\begin{equation}
v_m/v = (1/2-n)/(1/2-n_{m0}),
\end{equation}
The metallic phase would occupy the whole sample when the
total electron density $n$ is less than
$n_{m0}$.

\section{An example: the phase separated state with
metallic droplets}

As we argued above, the system with the short-range repulsion (1)
is unstable toward phase separation for $n$ close but different
from 1/2. The long-range Coulomb forces would however prevent the
full phase separation into large regions containing all extra
holes and the pure $n = 1/2$ charge ordered region. We can avoid
this energy loss by forming, instead of one big metallic phase
with many electrons, the finite metallic clusters with smaller
number of them. The limiting case would be a set of spherical
droplets, each containing one electron. This state is similar to
magnetic polarons (``ferrons'') considered in the problem of phase
separation in doped magnetic insulators
\cite{Nagaev,Kagan,Khomskii1}.

We present below the estimate
for the characteristic parameters of these droplets. The main aim
of this treatment is to demonstrate that the state constructed
in such a way will have the energy lower than the energy of the
homogeneous state, even if we treat these droplets rather crudely
and do not optimize all their properties. In particular, we will
make the simplest assumption that the droplets have sharp
boundaries and that the charge ordered state outside these droplets
is not modified in their vicinity. This state can be treated as a
variational one: if we optimize the structure of the droplet
boundary, its energy would only decrease.

The energy (per unit volume) of the droplet state with the
concentration $n_d$ of droplets can be written in total analogy
with the ferron energy in the double exchange model (see
\cite{Kagan,Khomskii1}). This yields
\begin{equation}
E_{droplet}=-tn_d
\left(z-\frac{\pi^2 a^2}{R^2}\right) -
\frac{W^2}{6Vz} \left[1-n_d \frac{4}{3}
\pi \left(\frac{R}{a} \right)^3 \right].
\end{equation}
Here, $a$ is the lattice constant and $R$ is the droplet radius.
The first term in (21) corresponds to the gain in kinetic energy
of electron delocalization inside the metallic droplets, and the
second term describes the charge ordering energy
in the remaining insulating part of the sample.

Actually, one should include the surface energy contribution to
the total energy of the droplet. The surface energy should be
of the order of $W^2R^2/V$.  For large droplets, this contribution
is small compared to the term $\propto R^3$ in (21); it would also
be reduced for a ``soft'' droplet boundary. One can show that even
in the worst case of a small droplet (of the order of a few lattice
constants) with the sharp boundary, this contribution would not
exceed about 20 percent of the bulk contribution. That is why we will
ignore this term below.

Minimization of the energy in (21) with respect to $R$ gives
\begin{equation}
\frac{R}{a}\propto \left(\frac{V}{t} \right)^{1/5}.
\end{equation}

The critical concentration $n_{dc}$ corresponds to the
configuration where metallic droplets start to overlap,
i.e. where the volume of the CO phase ( the second term
in (21)) tends to zero. Hence,
\begin{equation} n_{dc}=\frac{3}{4\pi}
\left(\frac{a}{R} \right)^3
\propto \left(\frac{t}{V} \right)^{3/5}.
\end{equation}

By comparing (12) with (21), (22), we see that for the deviations
from the half-filling $0 < \delta \leq \delta_c=1/2-n_{dc}$ the
energy of the phase separated state is always lower than the
energy of the homogeneous charge ordered state. The energy of the
droplet state (21) with the radius given by (22) is also lower
than the energy of the fully phase separated state, obtained by
the Maxwell construction from the homogeneous metallic state (17).
Correspondingly, the critical concentration $n_{dc}$ (23) is
larger than $n_{m0}$~(18). There is no contradiction here: the
droplet state, which we constructed has electrons confined in
spheres of radius $R$, and even when these droplets start to
overlap at $n=n_{dc}$, occupying the whole sample, the electrons
in this state, by construction, are still confined within their
own spheres and avoid each other. In other words, in our droplet
state certain degree of charge-ordering correlations is still
present, decreasing the repulsion and hence the total energy.

Thus, the energy of a phase separated state with the droplets
corresponds to the global minima of the energy for all
$0<\delta \leq \delta_c$. This justifies our conclusion
about phase separation into charge ordered state with $n=1/2$
and a metallic state with small spherical droplets.

The situation here resembles that of partially filled strongly
interacting Hubbard model, with the CO state corresponding to an
antiferromagnetic state of the latter and with the
nearest-neighbor interaction $V$ playing the role of the Hubbard's
$U$. In both cases, the kinetic energy of doped carriers tends to
destroy this ``antiferro'' or charge ordering, first ``spoiling''
it in their vicinity and finally leading to the formation of the
metallic state (Nagaoka ferromagnetism). In the Hubbard model, we
also face the situation with phase separation at a small enough
doping \cite{Visscher}.

Note also that for $ n > 1/2$ the compressibility of the charge-
ordered state is again negative
$\frac{1}{\kappa}=\frac{d^2 E}{d \delta^2}= - \frac{4 W^2}{3 V z}<0$
and has the same value as for the case $n<1/2$.
As a result, it is more favorable again to create a phase-separated
state for these densities. However, as it was already mentioned, the
nature of the second phase with $n > 1/2$ is not quite clear at
present, and therefore we do not consider this case here.

\section{An extended model}

Now we can extend the model discussed in the previous sections by
taking into account the essential magnetic interactions. In
manganites, besides the conduction electrons in $e_g$ bands, there
exist also practically localized $t_{2g}$ electrons, which we now
include to our consideration. The corresponding Hamiltonian has
the form
\begin{equation}
\hat{H}=-t \sum_{<i,j>,\sigma}c^+_{i\sigma}
c_{j\sigma}+V\sum_{<i,j>} n_i n_j -J_H\sum_i {\bf S}_i \sigma_i
+J\sum_{<i,j>} {\bf S}_i{\bf S}_j -\mu \sum_i n_i,
\end{equation}
In comparison to (1), the additional terms here correspond to the
strong Hund-rule onsite coupling $J_H$ between localized spins
${\bf S}$ and the spins of conduction electrons $\sigma$, and a
relatively weak Heisenberg antiferromagnetic (AFM) exchange $J$
between neighboring local spins. In real manganites, the AFM
ordering of the CE type in the CO phase is determined not only by
the exchange of localized $t_{2g}$ electrons but to a large extent
by the charge- and orbitally-ordered $e_g$ electrons themselves.
For simplicity, we ignore this factor here and assume the
superexchange interaction to be the same both in the CO and in the
metallic phases.

It is physically reasonable to consider this model in the limit
\[
J_H S>V>W>JS^2.
\]
In the absence of the Coulomb term, this is exactly the
conventional double exchange model (see e.g. \cite{Nagaev,Kagan}).
Note that the absence of doubly occupied sites in (20) is guaranteed
by the large Hund's term. It also favors the metallicity in the system,
since the effective bandwidth in our problem depends upon the magnetic
order. Therefore, the estimate for the critical concentration changes
here in comparison to (23). Similar to \cite{Kagan} the metallic
droplets will be ferromagnetic (FM) due to the double exchange.
The energy of one such droplet has the form
\begin{eqnarray}
E&=&-t \left(z-\frac{\pi^2 a^2}{R^2}\right)
- \frac{W^2}{6Vz} \left[1-\frac{4}{3} \pi
\left(\frac{R}{a} \right)^3 \right] + \\ &+&zJS^2
\frac{4}{3}\pi \left(\frac{R}{a} \right)^3
-zJS^2 \left[1-\frac{4}{3} \pi
\left(\frac{R}{a} \right)^3 \right].\nonumber
\end{eqnarray}

The last two terms in (25) describe respectively the loss
in the energy of the Heisenberg AFM exchange inside the FM metallic
droplets and the gain of this energy in the AFM insulating
part of the sample. The minimization with respect to the
droplet radius (as in (21)) yields
\begin{equation} \frac{R}{a} \propto \left( \frac{t}{V}
+\frac{JS^2}{t} \right)^{-1/5}.
\end{equation}

Note that at $t/V \ll JS^2/t$, formula (26) gives
just the same estimate for the radius of FM metallic
droplet $(R/a)\sim (t/JS^2)^{1/5}$ as in \cite{Nagaev,Kagan}.

In the opposite limit when $(t/V)\gg JS^2/t$, we reproduce the
same result $(R/a) \sim (V/t)^{1/5}$ as in (22). Finally, critical
concentration $n_c$ is estimated as follows
\begin{equation}
n_c \propto  \left( \frac{t}{V} +\frac{JS^2}{t}
\right)^{3/5}.
\end{equation}

As a result, taking into account also the tendency to the phase
separation at very small values of $n$
\cite{Nagaev,Moreo,Arovas,Khomskii1,Kagan} we come to the
following phase diagram for the extended model
(cf.~\cite{Khomskii1}:

\begin{enumerate}
\item At  $0<n<\left(\frac{JS^2}{t} \right)^{3/5}$,
it corresponds to the phase separation into a FM metal with $n
= n'> 0$ embedded in the AFM insulating matrix ($n = 0$). To minimize the
Coulomb energy, it may be again favorable to split this metallic region
into droplets with the concentration $n'$ and an average radius given
by Eq.~(26) with $t/V=0$, each containing one electron and kept apart
from one another.

\item At $\left(\frac{JS^2}{t} \right)^{3/5}<n<\left( \frac{t}{V}
+\frac{JS^2}{t} \right)^{3/5}<1/2$,
the system is a FM metal.
Of course, we need a window of parameters to satisfy the
inequality in the right-hand side. In actual manganites where
$t/V \sim 1/2 \div 1/3$
and $J/t \sim 1/3$, these conditions upon $n$ are not
necessarily satisfied. Experiments suggest that this window
is present for La$_{1-x}$Ca$_x$MnO$_3$, but it is definitely
absent for Pr$_{1-x}$Ca$_x$MnO$_3$ \cite{Khomskii1};

\item Finally, at $\left( \frac{t}{V} +\frac{JS^2}{t}
\right)^{3/5}<n<\frac{1}{2}$,
we have the phase separation
in the form of FM metallic droplets inside
the AFM charge ordered matrix.
\end{enumerate}

This phase diagram is in a good qualitative agreement with many
available experimental results for real manganites
\cite{Babushkina}--\cite{Voloshin}, in particular with
the observation of the small-scale phase separation
close to 0.5 doping \cite{Moritomo}.
Note also that our phase diagram has certain similarities with
the phase diagram obtained in \cite{Balents,Barzykin} for the
problem of spontaneous ferromagnetism in doped excitonic insulators.

\section{Conclusion}

Summarizing, we have shown that the narrow-band system, which
has the checkerboard charge ordering at $n = 1/2$ (corresponding
to the doping $x = 0.5$) is unstable toward phase separation
away from half-filling ($n \neq 1/2$). It separates into the
regions with the ideal CO ($n = 1/2$) and the other regions, in which
extra electrons or holes are trapped. The simplest form of these
metallic regions could be spherical metallic droplets embedded
into the CO insulating matrix. Simple considerations allow us to
estimate the size of these droplets and the critical concentration,
or doping $x_c = 1/2 - \delta_c$, at which the metallic phase would
occupy the whole sample and the CO phase would disappear. The account
of the magnetic interactions does not change these
conclusions but somewhat modifies the characteristic parameters
of the metallic droplets.

The long-range Coulomb interaction may also modify the results,
but we do not expect any qualitative changes. For the realistic
values of parameters, the size of metallic droplets is still
microscopic (about 10--30~$\AA$), and the excess charge in them will
be rather small.

The obtained picture corresponds rather well to the known
properties of 3D and layered manganites close to (less than) half
doping, $x \leq 1/2$. Percolation picture of transport properties
emerging from this treatment is confirmed by the results reported
in \cite{Arulraj,Gor'kov,Babushkina,Allodi,Voloshin,Moritomo};
moreover the coexistence of ferromagnetic reflections and those of
the CE type magnetic structure typical of the CO state at $x =
0.5$ were observed by the neutron scattering \cite{Kayumoto}.
Thus, the general behavior of underdoped manganites ($x \leq 0.5$)
is in a good qualitative agreement with our results.

Our treatment leads to the same tendency to phase separation
(instability of the homogeneous CO phase) also for overdoped
regime, $x > 0.5$. What would be the second phase in this case,
is not yet clear. Therefore we did not
concentrate our attention on such a situation.

Our treatment is applicable also to other systems with the charge
ordering, such as cobaltites \cite{Moritomo1} and nickelates
\cite{Alonso}. It would be interesting to study them for charge
carrier concentrations different from the commensurate
``checkerboard'' one

A number of important problems still remain unsolved
(the origin of the ``in-phase'' ordering in perovskite manganites in
$c$-direction, the detailed description of inhomogeneous states in
overdoped regime $n > 1/2$, the behavior at finite temperatures).
Nevertheless, in spite of the introduced simplifications, our model
seems to capture the essential physics underlying the interplay
between phase separation and charge ordering in transition metal
oxides.

\section*{Acknowledgments}

We are grateful to N.M.~Plakida and M.S.~Mar'enko for stimulating
discussions. D.Kh. expresses gratitude to S.-W.~Cheong and
Y.~Moritomo for the discussions of the experimental aspects of the
problem. The work was supported by INTAS (grants 97--0963 and
97--11954) and by the Russian-Dutch Program for Scientific
Cooperation funded by the Netherlands Organization for Scientific
Research (NWO). M.Yu.K. acknowledges the support of the Russian
President Program (grant 96--15--9694). The work of D.Kh.\ was
also supported by the Netherlands Foundation for the Fundamental
Research of Matter (FOM) and by the European network OXSEN.


\begin{thebibliography}{99}
\bibitem{Verwey} {E. Verwey, Nature (London) {\bf 144} 327 (1939);
E. Verwey and P.W. Haayman, Physica {\bf 8}, 979 (1941).}
\bibitem{Khomskii0} {D.I. Khomskii, Preprint of the P.N. Lebedev Physics
Institute no. 105 (1969).}
\bibitem{Mutou} {T. Mutou and H. Kontani, Phys. Rev. Lett. {\bf 83},
3685 (1999).}
\bibitem{Khaliullin} {J. van den Brink, G. Khaliullin, and D. Khomskii,
Phys. Rev. Lett. {\bf 83}, 5118 (1999).}
\bibitem{Jackeli} {G. Jackeli, N.B. Perkins, and N.M. Plakida,
cond-mat/9910391; Phys. Rev. B (in press).}
\bibitem{Jirak}{Z. Jirak {\it et al.} J. Magn. Magn. Mater. {\bf 53},
153 (1985).}
\bibitem{Arulraj} {A. Arulraj {\it et al.}, Phys. Rev. B {\bf 56}
R8115.(1998); M. Uehara {\it et al.} Nature {\bf 399}, 560
(1999).}
\bibitem{Nagaev} {E.L. Nagaev, Usp. Fiz. Nauk {\bf 166}, 833
(1996) [Phys. - Uspekhi {\bf 39}, 781 (1996)].}
\bibitem{Moreo}{A. Moreo, S. Yunoki, and E. Dagotto, Science {\bf 283},
2034 (1999).}
\bibitem{Arovas}{D. Arovas and F. Guinea, Phys. Rev. B {\bf 58}, 9150
(1998).}
\bibitem{Khomskii1} {D.I. Khomskii, Physica B {\bf 280}, 325 (2000).}
\bibitem{Uhrig}{G.S.~Uhrig and R.~Vlamink, Phys. Rev. Lett. {\bf 71},
271 (1993).}
\bibitem{Mori}{S. Mori, C.H. Chen, and S.-W. Cheong, Nature (London)
{\bf 392}, 473 (1998).}
\bibitem{Kagan}{M.Yu. Kagan, D.I. Khomskii, and M.V. Mostovoy, Eur.
Phys. J. B {\bf 12}, 217 (1999).}
\bibitem{Gor'kov}{L.P. Gor'kov and V.Z. Kresin, JETP Letters {\bf 67},
985 (1998).}
\bibitem{Yunoki}{S. Yunoki, T. Hotta, and E. Dagotto, Phys. Rev. Lett.
{\bf 84}, 3714 (2000).}
\bibitem{Soloviev}{I.V. Soloviev and K. Terakura, Phys. Rev. Lett.
{\bf 83}, 2825 (1999).}
\bibitem{Moritomo1}{Y. Moritomo, M. Takeo, X.J. Liu, T. Akimoto,
and A.Nakamura, Phys. Rev. B {\bf 58}, R13334 (1998).}
\bibitem{Alonso}{J.A. Alonso, J.L. Garc\'{i}a-Mu\~{n}oz, M.T.
Fern\'{a}ndez-D\'{i}az, M.A.G. Aranda, M.J. Mart\'{i}nez-Lope,
and M.T. Casais, Phys. Rev. Lett. {\bf 82}, 3871 (1999).}
\bibitem{Bulla} {P. Pietig, R. Bulla, and S. Blawid, Phys. Rev. Lett. {\bf
82}, 4046 (1999).}
\bibitem{Visscher} {P.B. Visscher, Phys. Rev. B {\bf 10}, 943 (1974).}
\bibitem{Babushkina} {N.A. Babushkina {\it et al.}, J. Phys.: Condens.
Matter {\bf 11}, 5865 (1999).}
\bibitem{Hennion} {M. Hennion {\it et al.}, Phys. Rev. Lett. {\bf 81},
1957(1998).}
\bibitem{Allodi} {G. Allodi, R. De Renzi, G. Guidi {\it et al.},
Phys. Rev. B {\bf 56}, 6036 (1997).}
\bibitem{Voloshin} {I.F. Voloshin {\it et al.},  Pis'ma v ZhETF {\bf 71},
157 (2000) [JETP Letters {\bf 71}, 106 (2000)].}
\bibitem{Moritomo} {Y. Moritomo, A. Machidas, S. Mori, N. Yamamoto, and
A.Nakamura, Phys. Rev. B {\bf 60}, 9220 (1999).}
\bibitem{Kayumoto} {R. Kayumoto, H. Yoshizawa, H. Kawano, Y. Tokura,
K. Ohoyama, and M. Ohashi, Phys. Rev. B {\bf 60}, 9506 (1999).}
\bibitem{Balents} {L. Balents and C.M. Varma, Phys. Rev. Lett. {\bf 84},
1264 (2000).}
\bibitem{Barzykin} {V. Barzykin and L.P. Gor'kov, cond-mat/9906401.}
\end{thebibliography}
\end{document}